\documentclass[12pt, a4paper]{article}

\usepackage{amsmath,amssymb,amsfonts}
\usepackage{graphicx}
\usepackage[colorlinks,hypertexnames=false]{hyperref}
\usepackage{cite}
\usepackage{appendix}
\usepackage{xcolor}
\definecolor{OG}{rgb}{0,0.6,0}
\definecolor{darkgreen}{rgb}{0,0.5,0.5}

\newcommand{\revisionC}[1]{\textcolor{darkgreen}{#1}}

\renewcommand{\vec}[1]{\mathbf{#1}}

\begin{document}

\title{Damping of dynamical friction force in  self-interacting ultralight dark matter and Fornax timing problem}
\author{E.V.~Gorbar${}^{1,2}$, O.V. Barabash${}^{1}$, V.M.~Gorkavenko${}^{1,2}$\thanks{Corresponding author. Email address: \textit{gorkavol@knu.ua}}, K. Korshynska${}^{3,4}$, \\
A.I. Momot${}^{1}$, 
A.O. Zaporozhchenko${}^1$\\
${}^1$ \it \small Faculty of Physics, Taras Shevchenko National University of Kyiv,\\
\it \small 64, Volodymyrs'ka str., Kyiv 01601, Ukraine\\
${}^2$ \it \small Bogolyubov Institute for Theoretical Physics, National Academy of Sciences of Ukraine,\\
\it \small 14-b, Metrolohichna str., Kyiv 03143, Ukraine\\
${}^3$ \it \small Institut für Mathematische Physik, Technische Universität Braunschweig,\\
\it \small Mendelssohnstraße 3, 38106 Braunschweig, Germany\\
${}^4$ \it \small Fundamentale Physik für Metrologie FPM, Physikalisch-Technische Bundesanstalt PTB,\\
\it \small Bundesallee 100, 38116 Braunschweig, Germany}
\date{}

\maketitle
\setcounter{equation}{0}
\setcounter{page}{1}%

\begin{abstract}

The dynamics of globular clusters in the Fornax dwarf galaxy pose a challenge for the standard cold dark matter and can be used to test other models of dark matter. We study this dynamics in the context of an ultralight bosonic dark matter model, accounting for the damping term in a generalized Gross-Pitaevskii equation. Employing analytic formulas for the dynamical friction force, the infall time and evolution of globular clusters are compared in the cases with and without the damping term. It is argued that the damping term plays an important role in the Fornax timing problem in ultralight dark matter (ULDM) models.  {
We found that the ULDM model with repulsive self-interaction can solve the Fornax timing problem in the absence of or with very small self-interaction, even if the initial position of the globular cluster is not far from the center of the galaxy. Still, the problem is resolved for strongly interacting repulsive ULDM, even for the most pressing case of globular cluster GC3, if its starting position exceeds 1.5 kpc.}

Keywords: {ultralight dark matter, Fornax dwarf galaxy, globular clusters, dynamical
friction force
}
\end{abstract}

\section{Introduction}\vspace{-0.5em}

Globular clusters are large and dense agglomerates of stars with typical half-light radius 3-5 pc up to tens of pc with masses $10^3M_\odot-10^6 M_\odot$  \cite{Gratton}. For example, it is estimated that the Milky Way contains more than 150 globular clusters with majority of them in the Milky Way halo. Globular clusters are observed also in dwarf galaxies, which typically orbit or interact with larger more massive galaxies like the Milky Way.

It was argued a long time ago that the dynamical friction can strongly affect the orbits of globular clusters \cite{Tremaine:1975}. One of the most studied cases is the Fornax dwarf spheroidal, which is a satellite of the Milky Way and contains six globular clusters (GC) \cite{Pace}. 
The orbital decay timescale for these globular clusters to sink toward the galactic nucleus due to dynamical friction is $\sim 1$ Gyr in the cold dark matter (CDM) model \cite{Tremaine:1976,Cole}, which is much less than the Hubble time.
 The Navarro–Frenk–White (NFW) profile \cite{Navarro:1996gj} of CDM has a cusp in the center of the galaxy. In such cuspy halos, dynamical friction acting on massive globular clusters becomes increasingly efficient as the cluster moves inward because the local dark matter density grows toward the center as $\sim 1/r$. As a consequence, standard estimates predict orbital decay times of order 1 Gyr for several of the observed Fornax globular clusters, significantly shorter than the age of the system. This constitutes the well-known “Fornax timing problem”: the clusters are expected to spiral into the galactic center and form a nuclear star cluster, whereas they are still observed today at kpc-scale distances.  However, if one chooses an initial position of GC at large distances ($r\gtrsim 1$ kpc) from the galactic center, where the CDM density in the NFW profile is small, then the dynamical friction force determined by the Chandrasekhar formula \cite{Chandrasekhar,Bar:2022liw}
is weak. Therefore, by choosing a sufficiently large initial orbital radius (for example, 1.5 kpc) for the globular cluster GC3, we obtain that the infall time of GC3 significantly exceeds the required time of 12 Gyr.

 It is worth mentioning also that this Fornax timing problem \cite{Oh} was revisited in \cite{Meadows} via N-body simulations, where it was concluded that dismissing the presence of a cusp in Fornax based on the spatial distribution of its GC population is unwarranted. Globular clusters were studied too in other dwarf galaxies. For example, recent study in \cite{Bar:2022liw} showed that the observed distribution of globular clusters in ultra-diffusive NGC5846-UDG1, harbouring around 59 globular clusters is indicative of their mass segregation induced by dynamical friction.

 One possible resolution to the Fornax globular cluster problem is the proposition that the dark matter halo in Fornax is cored \cite{Oh,Hernandez,Cole,Goerdt} with the core size of $r_c \geq 0.5$ kpc \cite{10.1093/mnras/stz573}. Then the Fornax timing problem is resolved due to ‘dynamical buoyancy’ created by the core. Dark matter distribution in massive and dwarf galaxies is naturally cored in models of dark matter composed of ultralight bosons with typical value of mass $10^{-22}\,\mbox{eV}$, where core is formed by the Bose-Einstein condensate (BEC) of ultralight bosons (for a review, see \cite{Ferreira:2020fam,Chavanis:2025qcg}). 
 
  {In ultralight dark matter (ULDM) models governed by the Gross–Pitaevskii (GP) equation, the central region of the halo develops a solitonic core with an approximately constant-density profile rather than a cusp as in the CDM model. Such cores substantially modify the efficiency of dynamical friction. In particular, the flattening of the density profile reduces the gravitational wake responsible for orbital decay. }
ULDM models have quite interesting phenomenology and are actively studied in the literature because while they could reproduce the large-scale structure of the Universe successfully explained in cold dark matter models, ULDM models are free of some problems which CDM models encounter at galactic scale. In addition, the ULDM self-interaction could also critically impact the dynamics and distribution of globular clusters in ultra-diffuse galaxies, leading to their segregation by mass \cite{Bar:2022liw}.

Further, dark matter halos in ultralight dark matter models have a nontrivial structure and consist of a solitonic core and an isothermal atmosphere \cite{chavanis2019predictive, Schive:2014hza}. The corresponding generalized Gross-Pitaevskii equation, which describes these halos, contains a logarithmic nonlinearity associated with a nonzero temperature and a source of dissipation. Employing the Madelung transformation, it is found \cite{chavanis2019predictive} that the latter enters the hydrodynamic equations as a damping term linear in hydrodynamic velocity. Heuristically, the presence of such a term could lead to the dissipation of perturbations in ultralight dark matter produced by a moving object and decrease the dynamical friction force, thus, alleviating the timing problem of Fornax globular clusters.

The orbital
decay timescale for the Fornax globular clusters to sink toward the galactic nucleus due to dynamical friction was considered, e.g., in \cite{Hui:2016ltb,Lancaster_2020} in different models of dark matter (DM). It was concluded that infall time could be greater than 10 Gyr for all globular clusters except GC3.   The case of globular cluster GC3 is the most problematic because GC3 is considered to be formed in situ and has an estimated age of approximately 12 Gyr \cite{deBoer2016_FornaxGCs,Bar:2021jff}.

Heuristic considerations mentioned above motivate us to study in this paper the impact of the damping term on the dynamical friction force acting on globular clusters moving in self-interacting ULDM environment and check whether it could help to resolve the Fornax timing problem, especially for the GC3 globular cluster.

The paper is organized as follows. The model used in our study is described in Sec.~\ref{sec:model}. The dynamical friction force acting on moving globular clusters is considered in  Sec.~\ref{sec:friction-force}. In Sec.~\ref{Sec: DM density in Fornax dwarf galaxy} the ultralight dark matter density in Fornax dwarf is described. In Sec.~\ref{sec:numerical-results} we compute the dynamical friction force in the Fornax dwarf galaxy.  Numerical results for the trajectory and infall time of moving globular clusters in the Fornax dwarf galaxy are given in Sec.~\ref{sec:infall}. Conclusions are drawn in Sec.~\ref{sec:Conclusion}.

\section{Model}
\label{sec:model}

Let us begin our analysis by writing down the generalized Gross-Pitaevskii equation \cite{chavanis2019predictive} which governs the\! dynamical evolution of self-gravitating\! and self-interacting\! BEC field\! $\psi$
\begin{equation} 
i\hbar\frac{\partial\psi}{\partial t} = \left(-\frac{\hbar^{2}}{2m}\nabla^{2} + m\Phi_{g}+ gN|\psi|^{2}+2k_BT\ln|\psi|\psi-\frac{i\hbar\xi}{2}\left[ \ln\left(\frac{\psi}{\psi^*}\right)-\ln\left(\frac{\psi^*}{\psi}\right)\right]\right)\psi,
  \label{GPP 1}
\end{equation}
where $g = {4 \pi \hbar^{2}a_{s}}/{m}$ is the coupling strength of the self-interaction of dark matter particles with mass $m$, and the $s$-wave scattering length $a_{s}$. Here $N$ is the number of boson particles, and $\hbar$ is the Planck constant. The first term on the right-hand side of the Gross-Pitaevskii equation (\ref{GPP 1}) corresponds to the kinetic energy. Repulsive self-interaction is described by the nonlinear term in Eq.~(\ref{GPP 1}) with $g>0$.  The fourth term accounts for an isothermal halo with an effective temperature $T$ surrounding the core. The last term is a damping term which ensures that the system relaxes to equilibrium, and its effect on the dynamical friction force is of principal interest to us in this study. The value of $\xi$ was estimated in \cite{chavanis2019predictive} via a generalized Einstein relation that expresses the fluctuation-dissipation theorem
\begin{equation}
\label{xi}
\xi=\frac{2k_BT}{\hbar},    
\end{equation}
where $k_B$ is the Boltzmann constant.

The second term in the generalized Gross-Pitaevskii equation (\ref{GPP 1}) contains the self-consistent gravitational potential $\Phi_g$, which supports the formation of a localized BEC core and is defined by the Poisson equation
\begin{equation}
  \nabla^{2}\Phi_{g} = 4\pi GmN|\psi|^{2},
  \label{GPP 2}
\end{equation}
where $G$ is the gravitational constant.

In our derivation of the dynamical friction force, we follow the setup developed in Refs.~\cite{Desjacques_2022,Buehler:2022tmr} where the dynamical friction force was given as double sum over the azimuthal and quantum numbers (for extension to the case of a Plummer sphere moving on a circular orbit, see Refs. \cite{Gorkavenko:2024upe,Barabash:2025ylw}) and generalize it to the case of nonzero $\xi$.
An orbiting object (star, globular cluster, dwarf galaxy, etc.) moving in a homogeneous ultralight dark matter composed of ultralight bosonic particles of mass $m$ perturbs the ULDM density $\rho(t,\mathbf{r})=\rho_\text{DM}(1+\alpha(t,\mathbf{r}))$ due to gravitational interaction. In the linear response approach, we find that the corresponding density inhomogeneity $\alpha(t,\mathbf{r})$ is governed by the equation
\begin{equation}
\partial_t^2\alpha-c^2_s\nabla_{\mathbf{r}}^2\alpha+\frac{\nabla_{\mathbf{r}}^4\alpha}{4m^2}+\xi\partial_t\alpha=4\pi G n(\mathbf{r}-\mathbf{r}_0(t)),
\label{perturbation-sphere}
\end{equation}
where $\mathbf{r}_0(t)$ denotes the position of the center of mass of the moving object and $n(\mathbf{r})$ is its mass density. In our study, we consider the case of a point perturber where $n(\mathbf{r})=M\delta^3(\mathbf{r})$ with object's mass $M$. Here $c_s$ is the sound velocity expressed through the ULDM pressure $P$ via the standard formula $c^2_s=\partial P/\partial \rho_\text{DM}$. In our case, the ULDM pressure is given by $P={\rho_\text{DM}^2 g}/{(2m^2)}+\rho_\text{DM} {k_B T}/{m}$ \cite{chavanis2019predictive} and therefore
\begin{equation}
\label{cs}
    c_s=\sqrt{\frac{\rho_\text{DM} g}{m^2}+ \frac{k_B T}{m}}.
\end{equation}

 {It should be noted that we consider the BEC regime of ULDM with $T\ll T_c$, where 
\begin{equation}
    T_c=\frac{2\pi \hbar^2}{m^{5/3}} \,\left(\frac{\rho_{DM}}{\zeta(3/2)}\right)^{2/3}
\end{equation}
 is the condensation temperature for an ideal Bose gas. This temperature $T_c\sim m^{-5/3}$ is very large for the boson mass in the ULDM model and is of order $10^{13}$ K for the boson particle mass $\sim 10^{-22}$ eV and $\rho_{DM}\sim10^{-21}$ $\textrm{kg/m}^3$. The case of $T>T_c$ was considered, e.g., in \cite{Hartman:2020fbg} for much heavier boson masses.}

In momentum space, Eq.(\ref{perturbation-sphere}) for steady-state motion gives
\begin{equation}
\alpha(\omega,\mathbf{k})=4\pi GM\!\!\int\limits^{+\infty}_{-\infty}\!\! dt'\,\frac{e^{i\omega t'-i\mathbf{k}\mathbf{r}_0(t')}}{-\omega^2+c^2_s k^2+\frac{k^4}{4m^2}-i\xi\omega}.
\label{perturbation-sphere-momentum-1}
\end{equation}
According to the Poisson equation, the density inhomogeneity $\alpha(t,\mathbf{r})$ and the moving object source a perturbation $\phi(t,\mathbf{r})$ in the gravitational potential
\begin{equation}   {\nabla}^2\phi(t,\mathbf{r})=4\pi G(\rho_\text{DM}\,\alpha(t,\mathbf{r})+M\delta^3(\mathbf{ r}-\mathbf{r}_0(t))).
\label{gravitational-potential}
\end{equation} 
Eq.~(\ref{gravitational-potential}) gives the following gravitational potential due to perturbed ULDM density (the complete perturbed gravitational potential is obviously the sum
$\phi=\phi_{\alpha}+\phi_0$, where $\phi_0$ is the Newtonian potential of the moving perturber):
\begin{equation}
\phi_{\alpha}(t,\mathbf{r})=-4\pi G \rho_\text{DM} \int \frac{d\omega d\mathbf{k}}{(2\pi)^4}\frac{\alpha(\omega,\mathbf{k})}{k^2}\,e^{-i\omega t+i\mathbf{k}\mathbf{r}},
\end{equation}
where $\alpha(\omega,\mathbf{k})$ is given by (\ref{perturbation-sphere-momentum-1}) and $\rho_\text{DM}$ can be approximated to be a constant in view of the hierarchy of masses of the Milky Way and perturber.

\section{Dynamical friction force with damping term}
\label{sec:friction-force}

Using the obtained gravitational potential of perturbed dark matter due to the moving perturber, we easily find the dynamical friction force
\begin{equation}
\mathbf{F}_\text{fr}(t)=-M\nabla_{\mathbf{r}}\phi_{\alpha}(t,\mathbf{r})_{|_{\mathbf{r}=\mathbf{r}_0(t)}}
=4\pi G \rho_\text{DM} M\int \frac{d\omega d\mathbf{k}}{(2\pi)^4}\frac{i\mathbf{k}}{k^2}\alpha(\omega,\mathbf{k})e^{-i\omega t+i\mathbf{k}\mathbf{r}_{0}(t)}.
\label{dynamical-force-local}
\end{equation}

It is convenient to change the variable $\tau=t-t'$ and then take into account that the integral over $\omega$ vanishes for $\tau<0$. Finally, taking into account Eq.~(\ref{perturbation-sphere-momentum-1}), we get the dynamical friction force
\begin{equation}
\mathbf{F}_\text{fr}(t)=\frac{G^2M^2\rho_\text{DM}}{\pi^2} \!\!\int\limits^{+\infty}_{0}\!\!d\tau \!\!\int\!\! d\omega d\mathbf{k}\frac{i\mathbf{k}}{k^2}\frac{e^{-i\omega\tau+i\mathbf{k}\mathbf{r}_0(t)-i\mathbf{k}\mathbf{r}_0(t-\tau))}}{-\omega^2-i\xi\omega+c^2_sk^2+\frac{k^4}{4m^2}}.
\label{dynamical-force-local-1}
\end{equation}
It is important to note that if the damping term is absent, then the same expression for the dynamical term applies with the replacement $\xi \to \delta$ where $\delta \to +0$ reproducing the dynamical friction force obtained in \cite{Berezhiani:2023vlo}.

The poles of the integrand in Eq.~(\ref{dynamical-force-local-1}) are given by
\begin{equation}
\omega_{1,2}=-\frac{i\xi}{2} \pm D(k), \quad D(k)=\sqrt{-\frac{\xi^2}{4}+c^2_sk^2+\frac{k^4}{4m^2}}.
\label{poles}
\end{equation}
Since they are always located in the lower half-plane of the complex plane $\omega$,  we should close the contour over $\omega$ in the lower half-plane for $\tau>0$. 
Integration over $\omega$ in Eq.~(\ref{dynamical-force-local-1}) yields
\begin{equation}
\mathbf{F}_\text{fr}(t)=\frac{2G^2M^2\rho_\text{DM}}{\pi}\!\int\limits^{+\infty}_{0}\!\! d\tau\!\int\!d\mathbf{k}\,\frac{i\mathbf{k}}{k^2D(k)}\,e^{i\mathbf{k}\mathbf{r}_0(t)-i\mathbf{k}\mathbf{r}_0(t-\tau)}e^{-\xi\tau/2}\sin\left(\tau D(k) \right).
\label{dynamical-force-sphere-total-2}
\end{equation}

Assuming that GC rotates on a circular orbit with angular velocity $\mathbf{\Omega}$ directed along the $z$-axis, we have 
$$
\mathbf{r}_0(t)=r_0(\cos(\Omega t),\sin(\Omega t),0),\quad \mathbf{k}=(\mathbf{k_p},k_z), \quad \mathbf{k_p}=(k_p\cos\phi,k_p\sin\phi)
$$
and
$$
\mathbf{k}\mathbf{r}_0(t)-\mathbf{k}\mathbf{r}_0(t-\tau)=k_pr_0\Big(\cos(\Omega t-\phi)-\cos(\Omega(t-\tau)-\phi)\Big),
$$
where $r_0$ is the orbit radius.

Setting without loss of generality $t=0$, we obtain 
\begin{multline}
\mathbf{F}_\text{fr}=\frac{2G^2M^2\rho_\text{DM}}{\pi}\!\!\int\limits^{+\infty}_{0}\!\! d\tau \!\!\int\limits^{+\infty}_{-\infty}\!\!dk_z\!\!\int\limits^{\infty}_0\!\! k_pdk_p\!\!\int\limits_{0}^{2\pi}\!\! d\phi\,\frac{i\mathbf{k}}{k^2D(k)}
 e^{ik_pr_0\big(\cos\phi-\cos(\phi+\Omega\tau)\big)}e^{-\xi\tau/2}\sin\left(\tau D(k)\right).
\label{dynamical-force-sphere-total-2.5}
\end{multline}
Obviously, the $z$-component of this force vanishes because the integrand is odd in $k_z$.

We can split the integral over $\phi$ in Eq.~(\ref{dynamical-force-sphere-total-2.5}) into two integrals, $\int_0^{2\pi}=\int_0^{\pi}+\int_\pi^{2\pi}$, and make the change of variable $\phi'=\phi -\pi$ in the second
\begin{equation}
    I_\phi=\int\limits_0^{2\pi}\!\!d\phi\, \mathbf{k}_p(\phi)    e^{if(\phi)}=\int\limits_0^{\pi}\!\!d\phi\,\mathbf{k}_p(\phi) e^{if(\phi)}+ \int\limits_0^{\pi}\!\!d\phi'\, \mathbf{k}_p(\phi'+\pi)e^{if(\phi'+\pi)},
\end{equation}
where $f(\phi)=k_pr_0\big(\cos\phi-\cos(\phi+\Omega\tau)\big)$ and $\mathbf{k_p}(\phi)=(k_p\cos\phi,k_p\sin\phi)$, hence $f(\phi'+\pi)=-f(\phi')$ and $\mathbf{k}_p(\phi'+\pi)=-\mathbf{k}_p(\phi')$. Thus, we get
\begin{equation}
I_\phi=\int\limits_0^{\pi}\!\!d\phi\,\mathbf{k}_p(\phi) \left(e^{if(\phi)}- e^{-if(\phi)}\right)=2i\!\!\int\limits_0^{\pi}\!\!d\phi\, \mathbf{k}_p(\phi)  \sin\left(f(\phi) \right).
\end{equation}
Using the obtained relation, we can express the dynamical friction force  
in the explicitly real form
\begin{multline}
    \mathbf{F}_\text{fr}= F_r \hat{\boldsymbol{r}} + F_t \hat{\boldsymbol{\phi}} =  -\frac{4G^2M^2\rho_\text{DM}}{\pi}\!\!\int\limits^{+\infty}_{0}\!\! d\tau \!\!\int\limits^{+\infty}_{-\infty}\!\! dk_z\!\!\int\limits^{\infty}_0\!\! k_pdk_p\!\!\int\limits^{\pi}_0\!\! d\phi\,\frac{\hat{\boldsymbol{r}} k_p\cos\phi + \hat{\boldsymbol{\phi}} k_p\sin\phi} {k^2D(k)}\\
    \times\sin\left[k_pr_0\big(\cos\phi-\cos(\phi+\Omega\tau)\big)\right]e^{-\xi\tau/2}\sin\left(\tau D(k)\right),
\label{dynamical-force-sphere-total-tangential-1}
\end{multline}
where \( F_r \) and \( F_t \) are the tangential and radial components of the dynamical friction force.

\section{DM density in Fornax  dwarf galaxy}
\label{Sec: DM density in Fornax  dwarf galaxy}

In what follows, we use the properties of DM halo of Fornax inferred from observations in \cite{jardel2012dark}, where the Fornax mass was modeled to be $M = 5.8^{+1}_{-0.2} \cdot 10^7 M_\odot$ as the best fit of the observations of stellar light profile and stellar kinematics. The authors also found that the DM halo density has to be approximately constant and of the order of $\rho_\textrm{obs} = (1.6 \pm 0.1) \cdot 10^{-2} M_\odot  \cdot \text{pc}^{-3}$. This observation thus disfavors a cuspy CDM density in the center and suggests the presence of a core, which is one of the prominent features of ULDM. For our analysis, we adopt the value of the density $\rho (r_h) = \rho_\textrm{obs}$, where $r_h \approx 710$~pc is the observed half-light radius of the Fornax halo \cite{mcconnachie2012observed}.  {We will use $r_h$ as an estimate of the ULDM core size and refer to it hereafter as the halo radius. While this set of observationally based constraints uniquely determines the model adopted in our analysis, we note that the structure and size of the DM Fornax halo remain under debate because the observed baryon kinematics can be fitted by multiple models, see Refs. \cite{walker2007velocity,Goldstein:2022pxu,Korshynska:2025nia}. }

Note that for a known density profile $\rho(r)$, one can easily derive the corresponding rotation velocity 
of a test particle (e.g., star) in the gravitational potential produced by $\rho(r)$ \cite{chavanis2011mass}
\begin{equation}
    v(r) = [G M(r)/r]^{1/2},\quad M(r) = 4\pi\!\! \int\limits_0^r\!\! dr' (r')^2 \rho (r'),
\end{equation}
where $M(r)$ is the total ULDM mass within radius $r$.

Having discussed the properties of the dark matter halo, we can now proceed to its description using ULDM models. A particular ULDM model is fixed when a specific boson candidate with mass $m$ and the s-wave scattering length $a_s$ is considered. Then, for given $\{m, a_s\}$, the ULDM halo density $\rho(r)$ can be found as a solution to Eq.~(\ref{GPP 1}). The density profile of this solution is dominated by the inner fully coherent core with slowly varying density \cite{chavanis2019predictive}. In the core region, the effective temperature term $\sim k_B T$ in Eq.~(\ref{GPP 1}) is small compared to the other terms and can be neglected \cite{chavanis2019predictive}, so that the core can be well-approximated by a solution to the ordinary GPP equations without dissipation. Beyond the core the effective temperature term dominates in the generalized GPP equation and gives rise to a dilute isothermal envelope. Thus, following the analysis in \cite{chavanis2019predictive}, we consider the approximate analytical expressions for the ULDM density $\rho(r)$ in these two regions separately.

We first focus on the central region of the ULDM halo, which is described by a BEC soliton solution of GPP equations \cite{chavanis2019predictive, chavanis2011mass}. Its shape is determined by dimensionless parameter $\chi$ \cite{chavanis2011mass}
\begin{equation}
    \chi = \frac{G M^2 m a_s}{\hbar^2},
\end{equation}
which sets the boundary between the Gaussian and Thomas-Fermi (TF) approximation regimes. The Gaussian ansatz approximates best the soliton density in the absence of short-range interactions ($a_s = 0$) and is valid only for small and moderate values of $\chi \lesssim 1$. In the other regime of $\chi \gg 1$, the repulsive interactions play a dominant role in counteracting gravitational attraction, leading to a TF density profile. While these density profiles will be discussed further in Secs.~\ref{Subsec: Thomas-Fermi density: strong repulsive interaction}-\ref{subsec: Gaussian density: Weak interaction}, we find that for Fornax
\begin{eqnarray}
    \chi = 1.42 \cdot 10^{83} \frac{a_s}{ \textrm{fm}} \frac{m}{\text{eV}}, \label{eq: chi parameter Fornax}
\end{eqnarray}
where ${\rm fm}=10^{-15}$ meters. 
This means that for a boson mass $\sim 10^{-22}$~eV, a characteristic value of the s-wave scattering length (for $\chi=1$) is of order $10^{-76}$ meters and decreases for larger boson masses.

\subsection{Thomas-Fermi density: strong repulsive interaction}
\label{Subsec: Thomas-Fermi density: strong repulsive interaction}

If the DM particle has a strong repulsive interaction, the TF approximation is applicable. In this approximation, the ground state density of the soliton reads
\begin{eqnarray}
    \rho_\text{TF} (r) =  \rho_0^\text{TF} \frac{\sin (\pi r/R_\text{TF})}{\pi r/R_\text{TF}}, \quad \rho_0^\text{TF}=\frac{\pi M}{4 R_\text{TF}^3}
    \label{eq: TF density}
\end{eqnarray}
and lies compactly within $r \in [0, R_\text{TF}]$. Fitting $\rho_\text{TF}(r_h) = \rho_\text{obs}$ fixes $R_\text{TF} = 1$~kpc and $\rho_0^\text{TF} = 4.56 \cdot 10^{-2} M_\odot/\text{pc}^3$.
Using the density profile (\ref{eq: TF density}), we can infer the properties of the ULDM. The TF radius is related to the ULDM parameters $\{m, a_s\}$ via 
\begin{equation}
    R_\text{TF} = \pi \left(\frac{a_s \hbar^2}{G m^3}\right)^{1/2}
\end{equation}
as shown in \cite{Chavanis}. This implies that
\begin{equation}
    \frac{a_s}{\textrm{fm}} \left(\frac{\text{eV}/c^2}{m}\right)^3 = 3.28\cdot 10^3,
    \label{eq: TF constraint}
\end{equation}
which is the same constraint as found in \cite{chavanis2019predictive}. In fact, this relation links the $s$-wave scattering length $a_s$ (or the coupling strength of the self-interaction $g = {4 \pi \hbar^{2}a_{s}}/{m}$) and the mass $m$ of a dark matter particle. Plugging this relation into \eqref{eq: chi parameter Fornax}, we obtain $m \gg 2.15 \cdot 10^{-22}$~eV. Therefore, we conclude that for boson masses $m > 10^{-21}$~eV, the ULDM density profile can be approximated by the TF density (\ref{eq: TF density}).

For the analytically found TF density distribution (\ref{eq: TF density}), one can infer the rotation curve \cite{chavanis2019predictive} 
\begin{eqnarray}
    v^2(r) &=& \frac{4 G \rho_0 R_\text{TF}^2}{\pi} \left[\frac{R_\text{TF}}{\pi r} \sin \left(\frac{\pi r}{R_\text{TF}}\right) - \cos \left(\frac{\pi r}{R_\text{TF}}\right)\right],\\
    \frac{v(r)}{\textrm{km}/\textrm{s}} &=& 15.8 \left[ \frac{318.3~\textrm{pc}}{r} \sin \left(\frac{r}{318.3~\textrm{pc}}\right) - \cos \left(\frac{r}{318.3~\textrm{pc}}\right) \right]^{1/2} \, ,
\end{eqnarray}
where the latter formula is given for the parameters of the Fornax ULDM halo defined above.

\subsection{Gaussian density: weak interaction}
\label{subsec: Gaussian density: Weak interaction}

\begin{figure}[t]
     \centering  \includegraphics[width=0.5\textwidth]{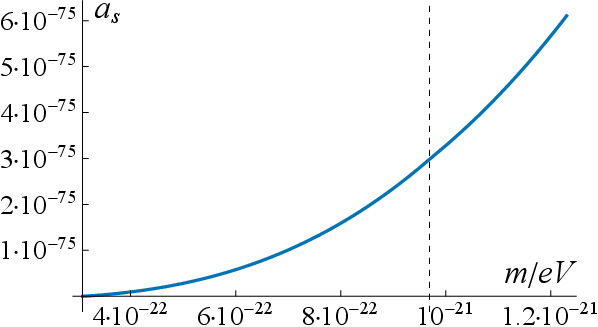}
   \caption{The dependence of $s$-wave scattering length $a_s$ (in meters) on the mass of the ULDM particle $m$ defined by Eq.~\eqref{eq: TF constraint} for $m> 10^{-21}$ eV and Eq.~\eqref{eq: first mass-radius relation} for $m<10^{-21}$ eV. The dashed vertical line separates the Gaussian density (left) from the TF (right) density approximation.}
   \label{Fig:asm}
 \end{figure}

In the case of non-interacting BEC or moderate interaction strength, one can approximate the ULDM density profile by the Gaussian ansatz \cite{chavanis2011mass}, which reads
\begin{equation}
    \rho_G (r) = \rho_0^G e^{-r^2/R^2},\quad \rho_0^G=\frac{M}{\pi^{3/2} R^3}.
    \label{eq: Gaussian Fornax density}
\end{equation}
The condition $\rho_\text{obs} = \rho_\text{G}(r_h)$ is best satisfied by the choice $R = 581$~pc in Eq.~(\ref{eq: Gaussian Fornax density}) giving the central density $\rho_0^G = 5.31 \cdot 10^{-2} M_\odot/\text{pc}^3$. We can again deduce the ultralight boson properties from the fixed ULDM halo properties. Imposing the condition that the soliton energy must be minimized, one can find the corresponding mass-radius relation \cite{chavanis2011mass,Chavanis:2019faf,Chavanis:2020rdo,Chavanis:2025qcg}
\begin{equation}
    R = \frac{\sigma}{\nu} \frac{\hbar^2}{G Mm^2} \left(1 \pm \sqrt{1 + \frac{6 \pi \zeta \nu}{\sigma^2} \frac{Gm M^2a_s}{\hbar^2}}\right),
    \label{eq: first mass-radius relationhbar}
\end{equation}
where $\sigma = 3/4$, $\nu = 1/\sqrt{2 \pi}$, and $\zeta = 1/(2 \pi)^{3/2}$. The sign '$+$' stands for the case of repulsive or no self-interaction $a_s \geq 0$, while the sign '$-$' represents the case of attractive self-interaction $a_s<0$. In what follows, we focus on the $a_s \geq 0$ regime. Plugging the known parameters $M$ and $R$ into this relation, we obtain the dependence of the $s$-wave scattering length $a_s$ (or the coupling strength of the self-interaction $g = {4 \pi \hbar^{2}a_{s}}/{m}$) on mass of the DM particle
\begin{equation}
    \left(\frac{m}{\text{eV}/c^2} \right)^2 = 4.76 \cdot 10^{-44} \left(1 + \sqrt{1 + 1.21 \cdot 10^{83} \frac{a_s}{\textrm{fm}} \frac{m}{\text{eV}}}\right)
    \label{eq: first mass-radius relation}
\end{equation}
which can be written in the form
\begin{equation}
    \frac{a_s}{\rm fm}=-3.47\cdot 10^{-40}\frac{m}{\rm eV}+3.65\cdot 10^{3}\left(\frac{m}{\rm eV}\right)^3.
\end{equation}
As one can see, in the non-interacting case $a_s = 0$, the above relation gives the boson mass $m = 3.09 \cdot 10^{-22}~\text{eV}$. For smaller masses $m$, ULDM self-interaction is attractive. The dependence of the $s$-wave scattering length $a_s$ on the mass of the DM particle $m$ is presented in Fig.\ref{Fig:asm}.

For the analytically determined Gaussian density (\ref{eq: Gaussian Fornax density}), one can derive the rotation curve \cite{chavanis2019mass} 
\begin{eqnarray}
    v^2(r) &=& \frac{GM}{R} \left[\frac{R}{r} \textrm{erf} \left(\frac{r}{R}\right) - \frac{2}{\sqrt{\pi}} e^{-(r/R)^2}\right],\\
    \frac{v(r)}{1 \textrm{km}/\textrm{s}} &=& 20.7  \left[ \frac{581~\textrm{pc}}{r} \textrm{erf} \left(\frac{r}{581~\textrm{pc}}\right) - \frac{2}{\sqrt{\pi}}\exp\left[-\left(\frac{r}{581~\textrm{pc}}\right)^2\right] \right]^{1/2}.
\end{eqnarray}

\subsection{Full ULDM halo with core and envelope}

Having described a compact soliton core in the central region, we now discuss the isothermal envelope on the outskirts of the ULDM halo. Its density has the shape of the NFW density profile \cite{Navarro:1996gj}, which is given by
\begin{equation}
    \rho_\text{NFW}(r) =  \frac{\rho_{e}}{r/r_e(1+r/r_e)^2},
    \label{eq: NFW envelope}
\end{equation}
while the gravitational potential and the mass enclosed within radius $r$ read
\begin{eqnarray}
    \Phi_\text{NFW}(r) &=& - \frac{4\pi G \rho_{e}r_e^3}{r}\ln\left(1 + \frac{r}{r_e}\right),\\
    M_\text{NFW}(r) &=& 4\pi\!\!\int\limits_{0}^{r}\!\! \rho(x)x^2 dx = 4\pi \rho_e r_e^3 \left[\ln\left(1 + \frac{r}{r_e}\right) - \frac{r/r_e}{1 + r/r_e}\right] \, .
\end{eqnarray}

To construct a realistic model of the Fornax dark matter halo, we need to account for the observations of the stellar velocity dispersion $\sigma_v$. The theory prediction can be found using \cite{binney2011galactic}
\begin{equation}
     \overline{v^2} = \frac{1}{\rho_\text{NFW}(r)} \int\limits_r^\infty\!\! dr' \rho_\text{NFW}(r') \frac{d\Phi_\text{NFW}}{dr'} 
\approx - \frac{\pi}{4} G \rho_{e} r_e^2 \frac{r_e}{r} \left[3 + 4 \ln\left(\frac{r_e}{r}\right)\right],
\end{equation}
where the last expression is given for $r_e/r \ll 1$ (far from the halo center). Thus, we find the velocity dispersion to be
\begin{equation}
    \sigma_{v} = (\overline{v^2})^{1/2} = \sqrt{G \rho_{e} r_e^2}\sqrt{- \frac{\pi}{4} \frac{r_e}{r} \left[3 + 4 \ln\left(\frac{r_e}{r}\right)\right]} \label{eq: veldispNFW} \, ,
\end{equation}
which is a slowly decaying function of $r$. On the outskirts of the BEC core, at distance $r = 1$~kpc (this is the TF radius $R_\textrm{TF}$ as found in Sec.~\ref{Subsec: Thomas-Fermi density: strong repulsive interaction}), according to the observations \cite{walker2007velocity}, the velocity dispersion of Fornax is $\sigma_{v} = 8$~km/s. At this distance Eq.~(\ref{eq: veldispNFW}) gives $\sigma_{v} = 0.32 \sqrt{G\rho_e r_e^2}$, where we used $r_e \approx 8$~pc (the exact value of $r_e$ will be determined in what follows). Then, we obtain the relation
\begin{equation}
   \rho_e r_e^2  = \frac{625 \text{ km}^2/\text{s}^2}{G} = 1.45 \cdot 10^5 M_\odot/\text{pc} \label{eq: kappa} \, 
\end{equation}
for parameters $\rho_e$ and $r_e$ characterizing the NFW envelope (\ref{eq: NFW envelope}).

\subsection{Soliton radius and continuous transition between soliton and envelope}

Following Ref.~\cite{Bar:2021jff}, we match the core (soliton) and envelope densities at a certain radial distance $r_t$ by fixing the corresponding densities $\rho_\text{NFW}(r_t) = \rho_s(r_t)$ and masses enclosed within radius $r_t$, namely, $M_\text{NFW}(r_t) = M_s(r_t)$.

If we model the soliton using the Gaussian density, see Sec.~\ref{subsec: Gaussian density: Weak interaction}, with
\begin{eqnarray}
    \rho_G(r)&=& \rho_0^{G} e^{-r^2/R^2},\\
    M_{s}(r)&=& \pi \rho_0^G R^2 \left[-2r e^{-r^2/R^2} + \sqrt{\pi}R\, \textrm{erf} \left(\frac{r}{R}\right)\right],
\end{eqnarray}
then the two continuity conditions at $r=r_t$ read
\begin{eqnarray*}
    \rho_{e} \frac{1}{r_t/r_e(1+r_t/r_e)^2} &=& \rho_0^G e^{-r_t^2/R^2},\\
   4\pi \rho_e r_e^3 \left[\ln\left(1 + \frac{r_t}{r_e}\right) - \frac{r_t/r_e}{1 + r_t/r_e}\right]  &=& \pi \rho_0 R^2 \left[-2r_t e^{-r_t^2/R^2} + \sqrt{\pi}R\, \textrm{erf} \left(\frac{r_t}{R}\right)\right]\, .
\end{eqnarray*}
Note that parameters $\rho_0^G  = 5.31 \cdot 10^{-2} M_\odot/\text{pc}^3$ and $R = 581$~pc are known, see Sec.~\ref{subsec: Gaussian density: Weak interaction}, and $\rho_e$ is derived from the observations of velocity dispersion, see Eq.~(\ref{eq: kappa}). Thus, the only unknown parameters are $r_e$ and $r_t$ which should be determined from the equations above.

For simplicity, we measure all masses in units of $M_\odot$ and all distances in units of kpc. Then we have the equations
\begin{eqnarray*}
    2.735 \frac{1}{r_e^2} \frac{1}{r_t/r_e(1+r_t/r_e)^2} &=&  e^{-r_t^2/0.581^2},\\
   10.94 r_e \left[\ln\left(1 + \frac{r_t}{r_e}\right) - \frac{r_t/r_e}{1 + r_t/r_e}\right]  &=&  0.581^3 \left[-3.442 r_t e^{-r_t^2/0.581^2} + \sqrt{\pi}\, \textrm{erf} \left(\frac{r_t}{0.581}\right)\right],
\end{eqnarray*}
giving $r_t = 1.238$~kpc and $r_e = r_t/165$. Thus, the transition between the core and envelope in this case occurs at distance $r_t$, where the density of both NFW and Gaussian soliton profiles equals $\rho(r_t) = 5.7 \cdot 10^{-4}~M_\odot$/\text{pc}$^3$. Finally, for $3.09\cdot 10^{-22}$~eV $<m<10^{-21}$~eV, we can write down the overall density of DM as
\begin{equation}
 \rho_\text{DM}(r) =    
\begin{cases}
 \rho_0^G e^{-r^2/R^2}, \,\, r \leq r_t\\    \rho_e \frac{1}{r/r_e (1 + r/r_e)^2},\,\, r> r_t ,
\end{cases}
\end{equation}
where $\rho_0^G = 5.31 \cdot 10^{-2} M_\odot/\text{pc}^3$, $\rho_e = 2.59 \cdot 10^3 M_\odot/\text{pc}^3$, $r_t = 1238$~pc, $R = 581$~pc, and $r_e = 7.50$~pc. 

For the mass of dark matter particles $m > 10^{-21}$~eV, the repulsive interaction is strong and the soliton should be modeled using the Thomas-Fermi model, see Sec.~\ref{Subsec: Thomas-Fermi density: strong repulsive interaction}, with the density
\begin{eqnarray}
    \rho_\text{TF}(r)&= &\rho_0^\text{TF} \frac{\sin (\pi r/R_\text{TF})}{\pi r/R_\text{TF}},\\
     M_{s}(r)&= &M \left[\frac{1}{\pi} \sin \left(\frac{\pi r}{R_\text{TF}}\right) - \frac{r}{R_\text{TF}} \cos \left(\frac{\pi r}{R_\text{TF}}\right)\right].
\end{eqnarray}

\begin{figure}[t]
    \centering  \includegraphics[width=\textwidth]{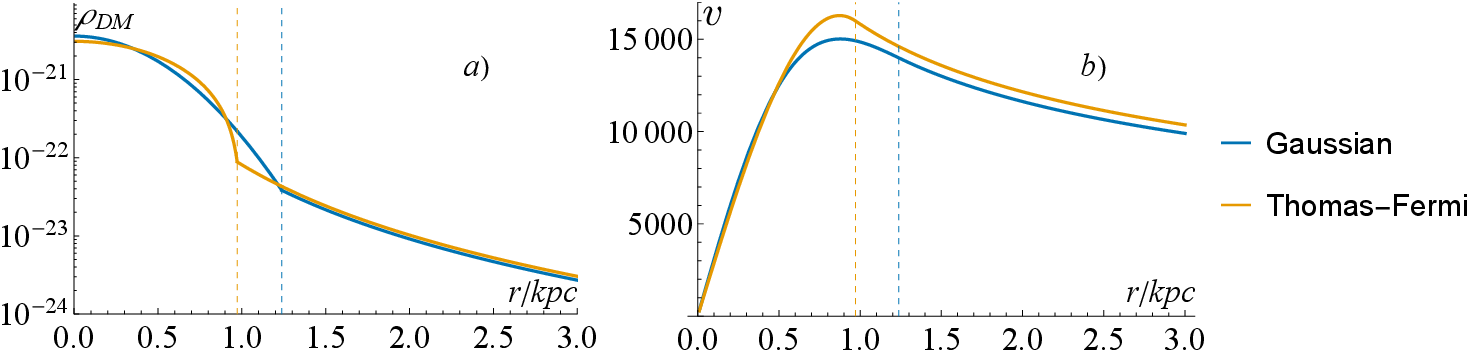} 
  \caption{Panel \textit{a}) Fornax mass density profiles $\rho_{DM}$ in  {the units of} $\textrm{kg/m}^3$. At the inflection points, the soliton dark matter density is stitched together with the NFW density profile. Panel \textit{b}) Fornax rotation curves  {with velocity $v$ in the units of m/s} for the TF and Gaussian density distributions of ULDM soliton.}
  \label{Fig: rot curves}
\end{figure}

Thus, at the transition radius $r = r_t$, we find the two continuity conditions
\begin{eqnarray*}
    \rho_{e} \frac{1}{r_t/r_e(1+r_t/r_e)^2} &=& \rho_0^\text{TF} \frac{\sin (\pi r_t/R_\text{TF})}{\pi r_t/R_\text{TF}},\\
   4\pi \rho_e r_e^3 \left[\ln\left(1 + \frac{r_t}{r_e}\right) - \frac{r_t/r_e}{1 + r_t/r_e}\right]  &=& M \left[\frac{1}{\pi} \sin \left(\frac{\pi r_t}{R_\text{TF}}\right) - \frac{r_t}{R_\text{TF}} \cos \left(\frac{\pi r_t}{R_\text{TF}}\right)\right].
\end{eqnarray*}
Here, parameters $R_\text{TF} = 1$~kpc and $M = 5.8 \cdot 10^7M_\odot$ are known and $\rho_e $ is given by \revisionC{(\ref{eq: kappa})}. To find the unknown $r_t$ and $r_e$, we reformulate the system of equations in dimensionless form, measuring masses in the units of $M_\odot$ and distances in kpc
\begin{eqnarray*}
     \frac{1}{r_e(1+r_t/r_e)^2} &=& 0.3136 \frac{\sin (\pi r_t)}{\pi }\, ,\\
   r_e \left[\ln\left(1 + \frac{r_t}{r_e}\right) - \frac{r_t/r_e}{1 + r_t/r_e}\right]  &=& 0.03177 \left[\frac{1}{\pi} \sin \left(\pi r_t\right) - r_t \cos \left(\pi r_t\right)\right],
\end{eqnarray*}
and find $r_e = 8.40$~pc and $r_t = 972.1$~pc. This transition radius is close to the $R_\text{TF}$ boundary of the soliton. Finally, for $m>10^{-21}$~eV, we can write down the overall density of ULDM as
\begin{equation}
 \rho_\text{DM}(r) =    
\begin{cases}
 \rho_0^\text{TF} \frac{\sin (\pi r/R_\text{TF})}{\pi r/R_\text{TF}}, \,\, r \leq r_t\\    \rho_e \frac{1}{r/r_e (1 + r/r_e)^2},\,\, r> r_t ,
\end{cases}
\end{equation}
where $\rho_0^\text{TF} = 0.0456$~$M_\odot$/\text{pc}$^3$, $\rho_e = 2.06 \cdot 10^3$~$M_\odot/$pc$^3$. 
The Fornax mass density profiles and rotation curves are presented in Fig.\ref{Fig: rot curves}.

\section{Numerical results for dynamical friction in Fornax}
\label{sec:numerical-results}

To analyse numerically the obtained dynamical friction force, it is convenient to introduce the following dimensionless variables $K_z=k_zr_0$, $K_p=k_p r_0$, and $\mathcal{T}=\tau c_s/r_0$. In addition, $\mathcal{M}=r_0\Omega/c_s$ is the Mach number, $A=mr_0c_s {/\hbar}$, and $\Xi=\xi r_0/c_s$. Then
\begin{equation}
  D(k)=\frac{\tilde{D}(K)}{2mr_0^2}, \quad \tilde{D}(K)=\sqrt{K^4+4A^2K^2-A^2\Xi^2} 
\end{equation}
and the tangential component of the friction force  (\ref{dynamical-force-sphere-total-tangential-1}) is
\begin{equation}\label{Ftdynforce}
 F_t=-\frac{4\pi G^2 M^2\rho_\text{DM}}{c_s^2} \mathcal{F}_t, 
\end{equation}
where
$$
\mathcal{F}_t=\frac{4A}{\pi^2 }\!\!\int\limits^{+\infty}_{0}\!\! d\mathcal{T} \!\!\int\limits^{+\infty}_0\!\! dK_z\!\!\int\limits^{\infty}_0\!\! K_pdK_p\!\!\int\limits^\pi_0\!\! d\phi\,\frac{K_p\sin\phi}{K^2\tilde{D}(k)}
$$
\begin{equation}
\times\sin\left[K_p\big(\cos\phi-\cos(\phi+\mathcal{M}\mathcal{T})\big)\right]e^{-\Xi \mathcal{T}/2}\sin\left(\frac{\mathcal{T}\tilde{D}(k)}{2A}\right)
\label{dynamical-force-sphere-total-radial-2}
\end{equation}
is the dimensionless tangential component of the dynamical friction force and we used the fact that the integrand is an even function of $k_z$. In our numerical calculations, we replace the upper limits of integration in (\ref{dynamical-force-sphere-total-radial-2}) with finite $\mathcal{T}_\text{max}$, $K_{z\,\text{max}}$, and $K_{p\,\text{max}}$ and then check that increasing these finite cut-offs does not affect the results.

It should be noted that $\xi$ and the coupling strength of the self-interaction $g$ are not independent parameters. 
The coupling strength of the self-interaction $g = {4 \pi \hbar^{2}a_{s}}/{m}$ is determined by Eqs.\eqref{eq: TF constraint} and \eqref{eq: first mass-radius relation}. The dependence of the s-wave scattering length $a_s$ on the mass of ULDM boson is shown in Fig.\ref{Fig:asm}.
While the value of $\xi$ is defined by Eq.~(\ref{xi}), the temperature of halo can be found from the condition \cite{chavanis2019predictive}
\begin{equation}
    k_B T=\frac{m \sigma_{v}^2}{2},
\end{equation}
where $\sigma_{v}  = 8$ km/s is the stellar velocity dispersion for the Fornax dwarf galaxy \cite{walker2007velocity}.

\begin{figure}
    \centering
    \includegraphics[width=1\textwidth]{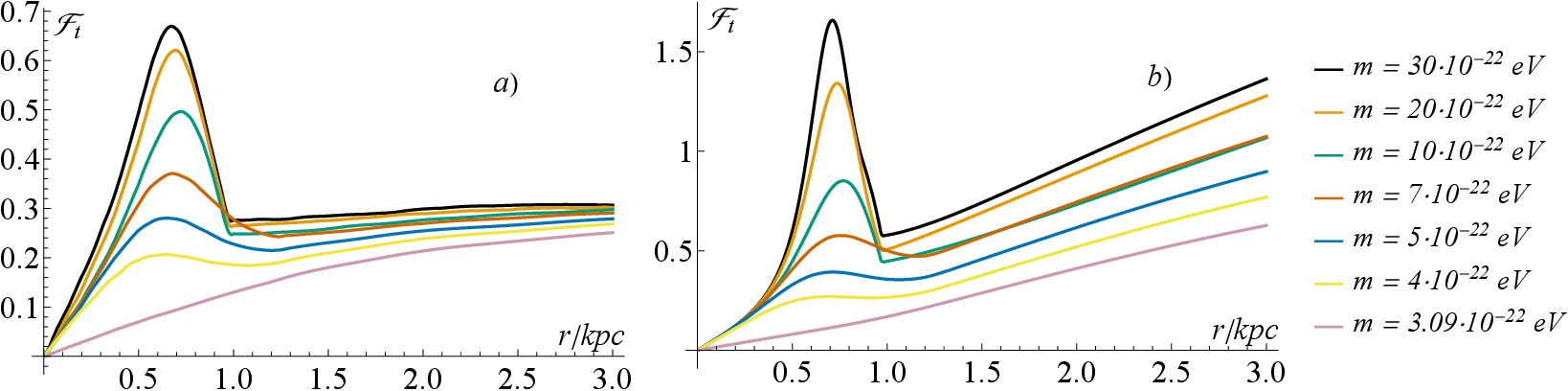}
    \caption{The tangential component of the dimensionless dynamical friction force $\mathcal{F}_t$ as a function of distance from the Fornax dwarf galaxy center for several values of the ULDM boson mass: panel \textit{a}) taking into account the damping term, panel \textit{b}) neglecting the damping term ($\xi=0$).} 
    \label{fig:Ft}
\end{figure}

Thus, we plot the tangential component of the dynamical friction force $\mathcal{F}_t$ as a function of only two parameters: the mass of ULDM boson $m$ and distance from the Fornax dwarf galaxy center,  see Fig.\ref{fig:Ft}. To compute the dynamical friction force for $m=3.09\cdot10^{-22}$~eV, $4\cdot10^{-22}$~eV,  $5\cdot10^{-22}$~eV, and $7\cdot10^{-22}$~eV, we used the Gaussian ansatz. For $m= 10^{-21}$~eV, $1.2\cdot10^{-21}$~eV, $2\cdot10^{-21}$~eV, and $2.7\cdot10^{-21}$~eV, the self-interaction is also repulsive and we use the Thomas-Fermi ansatz for the ULDM density.  It should be
noted that the lowest value of mass $m = 3.09 \cdot 10^{-22}$~eV is the minimum value for repulsive
ULDM self-interaction, which corresponds to the absence of self-interaction. For smaller
masses, the repulsive self-interaction turns into the attractive one.

According to Fig.\ref{fig:Ft}, in general, $\mathcal{F}_t$ increases as $m$ grows, and a quite strong dependence on the boson mass $m$ is observed for $r \leq 1$ kpc, resulting in a peak. 
The tangential component of the dynamical friction force attains a local maximum at $r\approx0.6$ kpc. However, for larger distances, the behaviour of $\mathcal{F}_t$ is different for the Gaussian and Thomas-Fermi regimes.

Comparing the results for the dynamical friction force in panels a) and b) of Fig.\ref{fig:Ft}, one can see that taking into account the damping term essentially modifies (decreases) the magnitude of the dynamical friction force as well as change its behavior at large distances.

\section{Trajectory and infall time of GC3}\label{sec:infall}

Having determined the dynamical friction force, we could now proceed to the evolution and motion of globular clusters. To describe them, we numerically solve the system of differential equations of motion
\begin{equation}\label{EqMot}
    \begin{cases}
        m_\text{GC}\ddot{r} = -\dfrac{G m_\text{GC} M(r)}{r^2}
        + \dfrac{l^2}{m_\text{GC}r^3}, \\[0.8em]
        \dfrac{dl}{dt} = -r(t)F_t(r),
    \end{cases}
\end{equation}
where $m_\text{GC}$ is the mass of GC, $l = m_\text{GC}r^2\dot{\phi}$ is the angular momentum, see details in Appendix.

\begin{figure}[t]
    \centering
   \includegraphics[width=\textwidth]{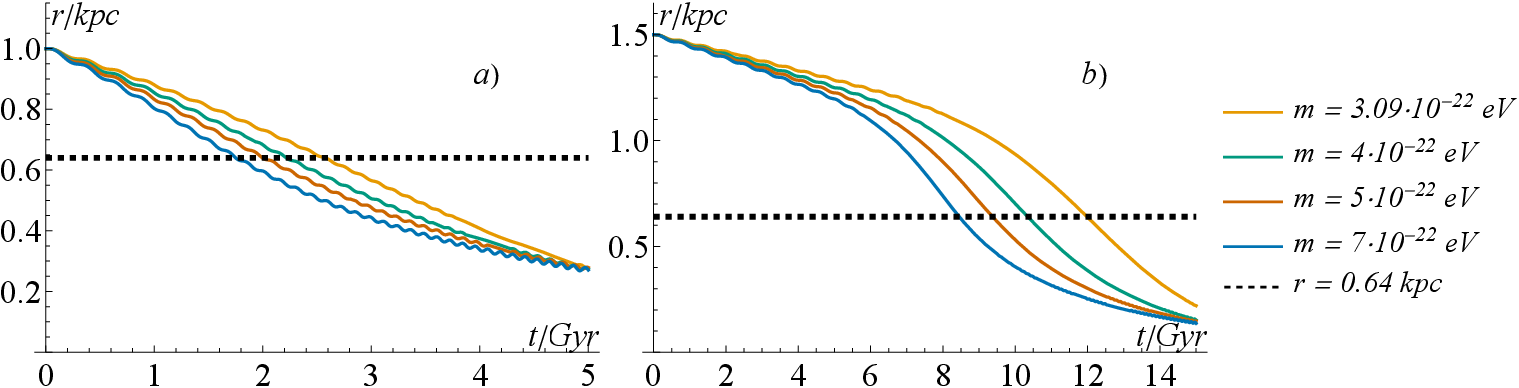}
    \caption{The trajectory $r(t)$ of the globular cluster GC3 as a function of time in the Gaussian regime and several values of the ULDM boson mass in the range $3.09\cdot 10^{-22}$~eV $<m<$ $7\cdot 10^{-22}$~eV with its initial position panel \textit{a}) $r_0=1$ kpc and panel \textit{b})  $r_0=1.5$ kpc.  The black horizontal line indicates the current position of GC3.}
    \label{fig:trajectG}
\end{figure}

\begin{figure}[b]
    \centering
   \includegraphics[width=\textwidth]{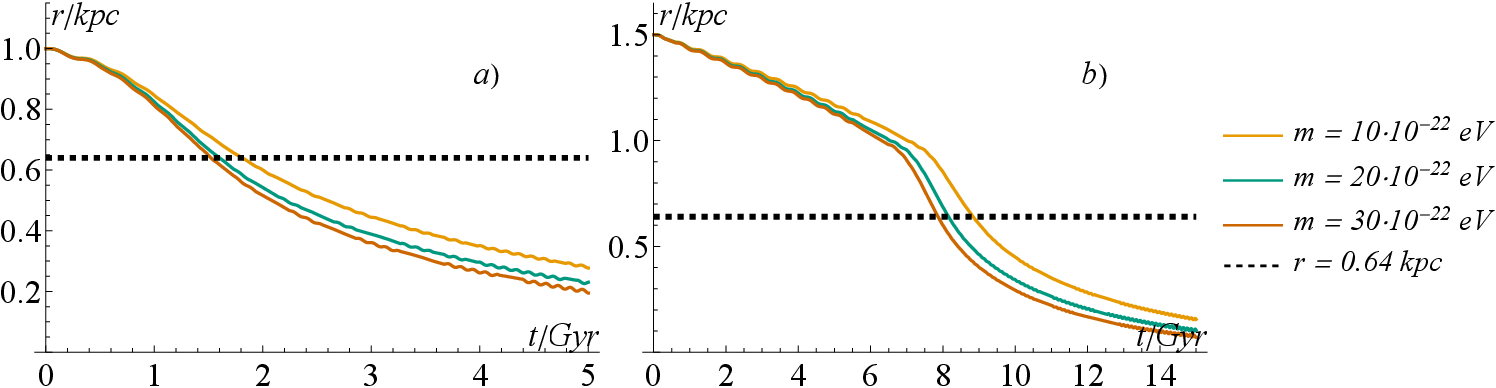}
    \caption{The trajectory $r(t)$ of the globular cluster GC3 as a function of time in the Thomas-Fermi regime for several values of the ULDM boson mass in the range $ 10^{-21}$~eV $<m<$ $3\cdot 10^{-21}$~eV with its initial position panel \textit{a}) $r_0=1$ kpc and panel \textit{b})  $r_0=1.5$ kpc.  The black horizontal line indicates the current position of GC3.}
    \label{fig:trajectTF}
\end{figure}

Setting the initial position of the globular cluster GC3 to be $r_0=1$ kpc and $r_0=1.5$ kpc, zero initial value of its radial velocity, assuming its mass to be $m_\text{GC}=4.98\cdot10^5M_\odot$ \cite{Bar:2021jff}, and solving the equations of motion \eqref{EqMot}, we plot the time dependence of $r(t)$ for the globular cluster GC3 in Fig.\ref{fig:trajectG} in the Gaussian regime  (for several values in the range $3.09\cdot 10^{-22}$~eV $<m<$ $7\cdot 10^{-22}$~eV) and Thomas-Fermi regime  in Fig.\ref{fig:trajectTF} (for several values in the range $10^{-21}$~eV $<m<$ $3\cdot 10^{-21}$~eV). In both regimes, we see that the infall time decreases as mass $m$ grows.

As one can see, in the Gaussian regime of weak repulsive self-interacting ULDM, which is valid for $3.09\cdot 10^{-22}$~eV $<m\lesssim$ $ 10^{-21}$~eV, we cannot obtain the required time of 12 Gyr for GC3 to move from its assumed initial position $r_0=1$ kpc from the center of the Fornax galaxy to its current position at $r_{GC3}=0.64$ kpc \cite{Bar:2021jff}. However, if we assume that GC3 was formed at larger distance, e.g., $r_0=1.5$ kpc, where the density of dark matter and the force of dynamic friction are smaller, then it is possible to explain the observed lifetime of the cluster provided that  $m\gtrsim  3.09\cdot 10^{-22}$~eV. This value corresponds with $a_\textrm{s} = 0$ and sets the boundary between the regimes of repulsive and attractive self-interaction. This means that, for the initial position $r_0=1.5$ kpc, the infall time of GC3 can be explained also for non-interacting (fuzzy) ULDM.

For $m\gtrsim 10^{-21}$~eV, the Thomas-Fermi regime of strongly repulsive self-interacting ULDM is realized. According to the results of Fig.\ref{fig:trajectTF}, one can see that the maximum time for GC3 to move from the initial position $r_0=1$ kpc from the center of the Fornax galaxy to its current position at $r_{GC3}=0.64$ kpc is only 2 Gyr, and for initial position $r_0=1.5$ kpc is about 9 Gyr for the mass $m=10^{-21}$ eV and decreases as the mass increases. Thus, in this case, we cannot provide the necessary infall time.  

\begin{figure}[t]
    \centering
   \includegraphics[width=0.45\textwidth]{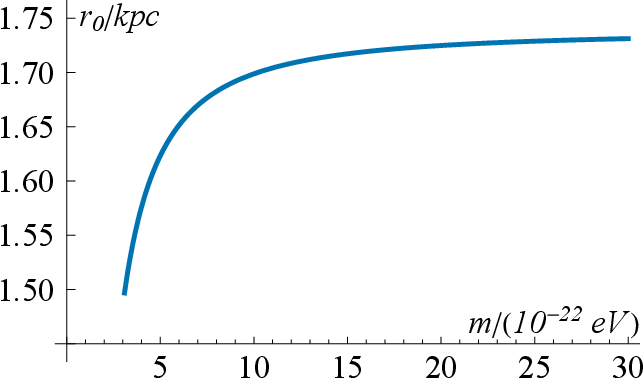} 
    \caption{The initial radial position $r_0$ of GC3 required to reach its current position 0.64 kpc as a function of the mass of ULDM boson.}
    \label{fig:initialposition}
\end{figure}

 {Since we do not know the initial position of GC3 and it is currently located 0.64 kpc from the center of the galaxy, it makes sense to fix the GC3’s current position and determine the initial position from which it reached its current position over 12 Gyr as a function of the mass of ULDM boson. The corresponding graph is shown in Fig.\ref{fig:initialposition}. One can see that the larger the mass of the ULDM boson, the greater the initial radial distance of GC3 from the center of the galaxy must be. In addition, it can be concluded that if the initial position of GC3 exceeds 1.5 kpc, then a sufficiently wide range of ULDM boson mass of order $10^{-21}$ eV is allowed. }

\begin{figure}[h]
    \centering
    \includegraphics[width=\textwidth]{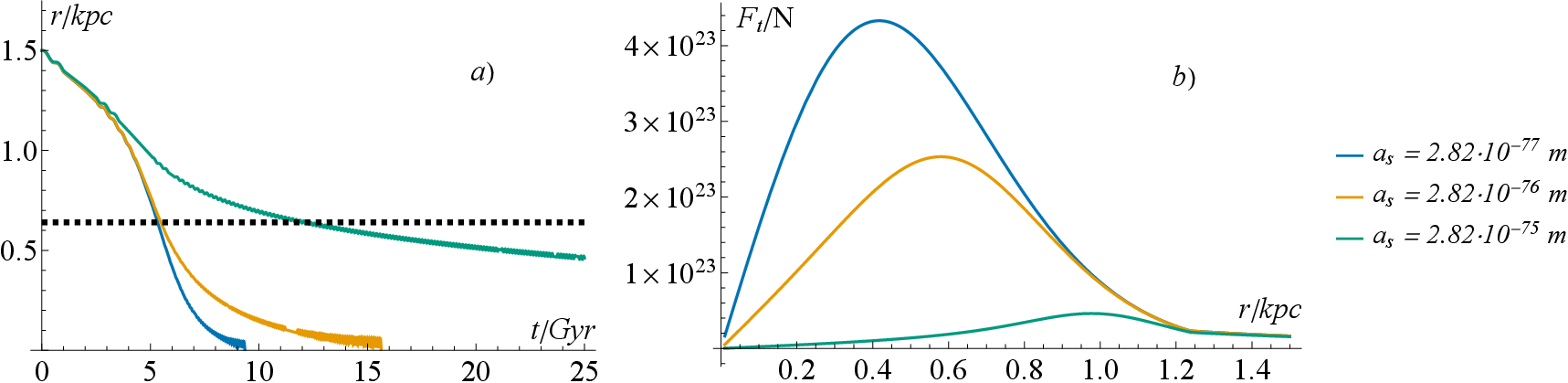}
    \caption{Panel \textit{a}) the trajectory $r(t)$ of the globular cluster GC3 as a function of time in the Gaussian regime for different values of repulsive coupling of self-interaction $g$.   The black horizontal line indicates the current position of GC3. Panel \textit{b}) the tangential component of the dynamical friction force $F_t$  {(in the units of N)} given by Eq.\eqref{Ftdynforce} as a function of distance to the center of the galaxy. In both panels, the  {yellow} line is plotted for coupling $g$ defined by the mass-radius relation \eqref{eq: first mass-radius relation} with $m=5\cdot10^{-22}$ eV.}
    \label{fig:lambda}
\end{figure}

 {In Ref.\cite{Koo:2025jkx}, it was shown that repulsive self-interaction decreases the dynamical friction. This result is intuitive, as the repulsive force would reduce the clumps and hence reduce the dynamical friction. 
At the first sight, our results for the dynamical friction force for repulsive self-interacting ULDM contradict the results obtained in \cite{Koo:2025jkx}. 
However, compared to the study in \cite{Koo:2025jkx}, we imposed an additional constraint on the boson particle mass and the coupling constant, which follows from the boson mass-radius relation and connects the boson particle mass and the radius of the soliton core.
Therefore, to show the consistency of our results with those obtained in \cite{Koo:2025jkx}, we fixed the dark matter particle mass and investigated the effect of varying the coupling constant on the dynamical friction force. We found, in agreement with the results obtained in \cite{Koo:2025jkx}, that increasing the coupling constant of the repulsive interaction indeed leads to a decrease in the dynamical friction force and, hence, the GC infall time increases. Our results are shown in Fig.\ref{fig:lambda}  at the fixed value of the boson mass $m=5 \cdot 10^{-22}$ eV for repulsive coupling of self-interaction $g$ defined by the mass-radius relation \eqref{eq: first mass-radius relation} and for coupling ten times larger and smaller.}

\begin{figure}[t]
    \centering
   \includegraphics[width=\textwidth]{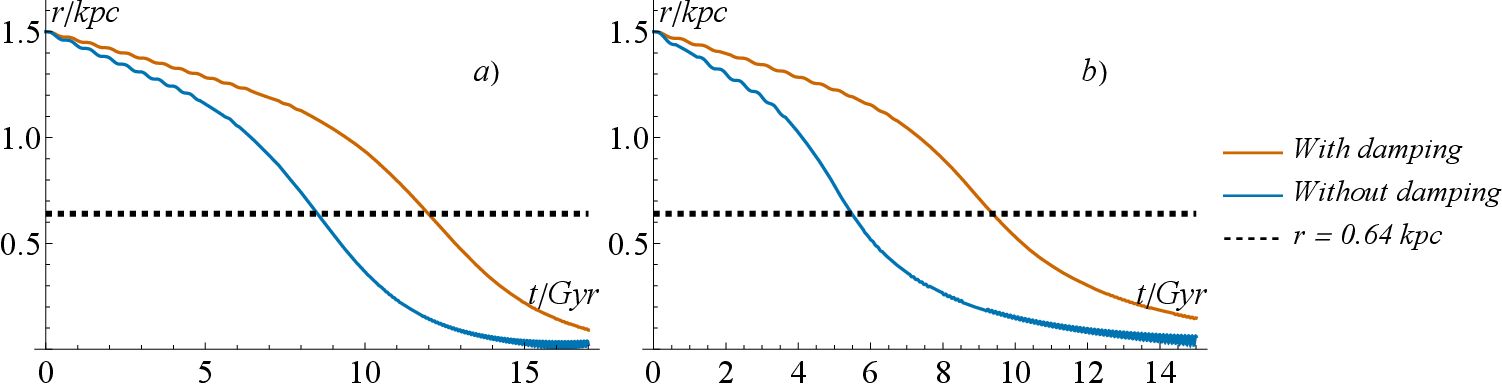}
    \caption{The trajectory $r(t)$ of the globular cluster GC3 as a function of time for several values of the ULDM boson mass  $m = 3.09\cdot 10^{-22}$~eV and $5\cdot 10^{-22}$~eV with its initial position  $r_0=1.5$ kpc in two cases with and without a damping term.  The black horizontal line indicates the current position of GC3.}
    \label{fig:damp_vs_no_damp}
\end{figure}

\begin{figure}[b]
    \centering
   \includegraphics[width=0.6\textwidth]{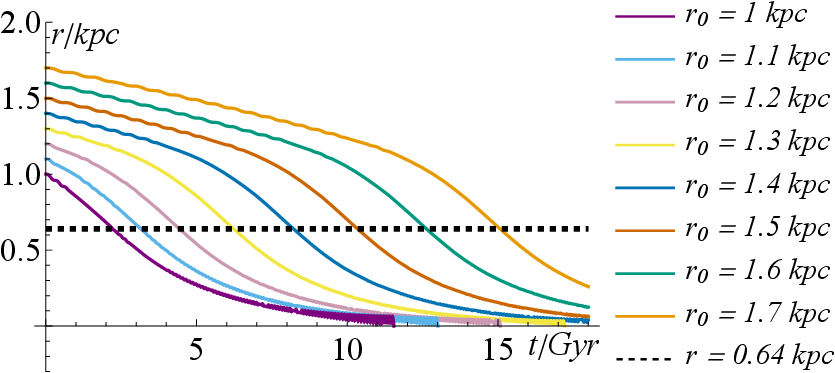}
    \caption{The trajectory $r(t)$ of the globular cluster GC3 as a function of time for the value of the ULDM boson mass  $m = 4\cdot 10^{-22}$~eV with its initial position in the range 1~kpc  $<r_0<1.7$ kpc.  The black horizontal line indicates the current position of GC3.}
    \label{fig:r0_dif}
\end{figure}

The role of the damping term ($\xi\neq0$) is demonstrated in Fig.\ref{fig:damp_vs_no_damp} for two values of mass $m = 3.09\cdot 10^{-22}$~eV and $5\cdot 10^{-22}$~eV. Clearly, damping results in an increase of the infall time. 
Fixing the DM mass $m=4\cdot 10^{-22}$ eV and including the damping term, the trajectory $r(t)$ of the globular cluster GC3 as a function of time is plotted in Fig.\ref{fig:r0_dif} for different initial positions in the range 1~kpc  $<r_0<1.7$ kpc.  The black horizontal line indicates the current position of GC3. As expected, the infall time grows as the initial distance increases.

\begin{figure}[t]
    \centering
    \includegraphics[width=\textwidth]{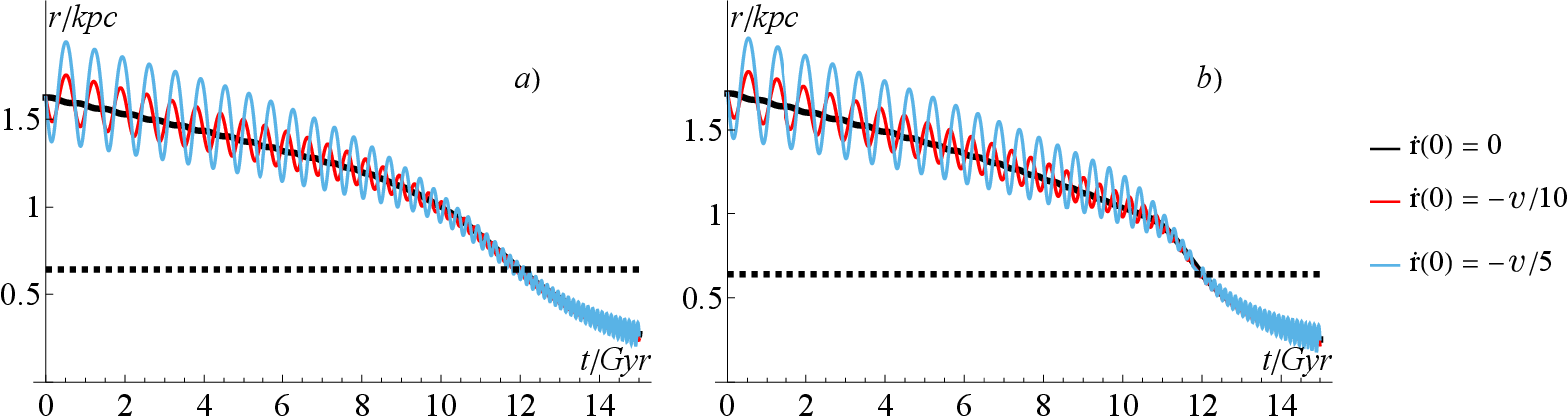}
    \caption{The trajectory $r(t)$ of the globular cluster GC3 as a function of time  for different initial radial velocity $\dot r(0)$, given in units of the initial tangential velocity $v$ derived from the rotation curve. Panels \textit{a}) and \textit{b}) correspond to ULDM boson masses  $m = 5\cdot 10^{-22}$~eV and $m = 2\cdot 10^{-21}$~eV, respectively.  The black horizontal line indicates the current position of GC3.}
    \label{fig:dotr0}
\end{figure}

 {In the present work, we considered only the case of circular orbits of GCs. However, the motion of globular clusters on elliptic orbits is also of considerable interest. In this case, when solving the differential equations of motion \eqref{EqMot}, one must take into account that the initial radial velocity of  globular cluster is nonzero, $\dot r(0)\neq 0$.
As an illustration, we determined the trajectory of GC3 moving in a ULDM halo with particle masses $m=5\times10^{-22}\,\mathrm{eV}$ and $m=2\times10^{-21}\,\mathrm{eV}$. Our computations, see Fig.\ref{fig:dotr0}, show that dynamical friction initially reduces the orbital eccentricity. However, as the cluster approaches the central region, the eccentricity increases again. Physically, dynamical friction tends to circularize initially elliptic orbits because energy losses are most efficient near the pericenter, where the ULDM density and the orbital velocity are highest. Consequently, the eccentricity decreases during the early stages of orbital decay. As the orbit shrinks and enters regions with a different density and potential structure, in particular, the central core, the relative rates of energy and angular momentum loss may change, allowing the eccentricity to grow again, see also \cite{Szolgyen2022}.}

 {
The actual eccentricity of the GC3 orbit remains unknown. Nevertheless, as can be seen from Fig.\ref{fig:dotr0}, at the current position of GC3 the orbital eccentricity tends to be close to zero over a fairly wide range of initial radial velocities.
}

\section{Conclusions}
\label{sec:Conclusion}

In this work, we investigated the dynamical friction force acting on orbiting globular clusters in the Fornax dwarf galaxy. We modeled dark matter composed of ultralight bosons by the generalized Gross-Pitaevskii equation and, in particular, studied the impact of the damping term. The case of the globular cluster GC3 is the most interesting and problematic because GC3 is considered to be formed in situ, has an estimated age of approximately 12 Gyr, and is located at present at a large distance of 0.64 kpc from the center of the Fornax dwarf galaxy.

We derived analytical expression for the dynamical friction force \eqref{dynamical-force-sphere-total-tangential-1} acting on circularly moving point objects in ULDM models both with and without the damping term in the generalized Gross-Pitaevskii equation \eqref{GPP 1}. This force arises due to gravitational interaction with the ULDM density perturbation induced by the moving object.

In our numerical calculations, we considered only the case of repulsive self-interacting ULDM in the regimes of weak (Gaussian) and strong (Thomas-Fermi) self-interaction. 
We found that there is a lower bound of ULDM boson mass $m_{low}\approx 3.09\cdot 10^{-22}$~eV. As one can see from the mass-radius relation \eqref{eq: first mass-radius relation}, for smaller values of mass, the repulsive interaction turns into an attractive interaction.

Results of the computation of the dynamical friction force as a function of ULDM boson mass and distance from the center of Fornax dwarf galaxy are given in Fig.\ref{fig:Ft} for the case with and without the damping term in the Gross-Pitaevskii equation. Our main finding is that the damping term essentially modifies (decreases) the magnitude of the dynamical friction force acting on stars, globular clusters, or dwarf galaxies moving in the ultralight dark matter halo. This decrease takes place because damping inhibits the ULDM density perturbation caused by moving objects. 

For the globular cluster GC3, we calculated its dynamical trajectory and found the infall time from the initial point at distances $r_0=1$ kpc and $r_0=1.5$ kpc to the current position at $r_{GC3}=0.64$ kpc from center of the Galaxy, see Fig.\ref{fig:trajectG} and Fig.\ref{fig:trajectTF}. According to our results, in the case of weakly self-interacting (the Gaussian regime) ULDM, the infall time of GC3 can attain the required value $T_{GC3}=12$ Gyr for a ULDM boson mass $m\gtrsim m_{low}$, provided that ULDM is non-interacting or has a very small repulsive self-interaction.  
In the strongly-interacting ULDM (the Thomas-Fermi regime), there is no solution for the infall time be equal to $T_{GC3}$ if the starting position is $r_0\lesssim 1.5$ kpc.  
 {However, if the initial position of GC3 exceeds 1.5 kpc, then a sufficiently wide range of ULDM boson masses of order $10^{-21}$ eV is allowed, see Fig.\ref{fig:initialposition}.}

To illustrate more clearly the impact of the damping term ($\xi \neq 0$), we plot the trajectory $r(t)$ of the globular cluster GC3 as a function of time in Fig.\ref{fig:damp_vs_no_damp} in two cases with and without a damping term. The inclusion of damping noticeably delays the orbital decay, leading to a significantly longer infall time compared to the case without damping.

To further explore this effect, we fix the dark matter mass at $m = 4 \cdot 10^{-22}$~eV and analyze the time evolution of the orbital radius $r(t)$ for the globular cluster GC3 as a function of initial distance. The corresponding trajectories are shown in Fig.\ref{fig:r0_dif} for a range of initial positions $1~\text{kpc} < r_0 < 1.7~\text{kpc}$. As expected, clusters starting at larger radii require more time to reach the observed current position.
For sufficiently large initial distances $r_0$, the infall time can match the required values even for higher dark matter masses $m$. In this case, viable solutions are not restricted to the Gaussian regime but can also be realized within the Thomas–Fermi regime.

 {Additionally, we considered the motion of GC3 on elliptic orbits and found that the orbital eccentricity is close to zero at the current position of GC3 for a fairly wide range of initial radial velocities, see Fig.~\ref{fig:dotr0}.}

 {Finally, we would like to note that our results rely on the input parameters derived from the analysis of astrophysical observations in \cite{Berezhiani:2023vlo,jardel2012dark,chavanis2019mass}, and we use the core-halo model to describe the density profile of the Fornax galaxy. For the ULDM model with the core size of $r_h \gtrsim  0.5$ kpc, our findings provide a robust estimate of the dynamical friction effect and may serve as a benchmark for further investigations.
}

\phantom{bjjbj}

\centerline{\bf Acknowledgements}
\vspace{5mm}

The authors are grateful
to B.I. Hnatyk and A.I. Yakimenko for fruitful discussions and helpful comments.  K.K. acknowledges funding by the Deutsche Forschungsgemeinschaft (DFG, German Research Foundation) under Germany’s Excellence Strategy -- EXC 2123 QuantumFrontiers -- 390837967. The work of E.V.G., V.M.G., and A.O.Z. was partially supported by the project 'Search for dark matter and particles beyond the Standard Model' of the Ministry of Education and Science of Ukraine (25BF051-01).

\newpage

\section*{Appendix}
\renewcommand{\theequation}{A.\arabic{equation}}
\setcounter{equation}{0}

The motion of a globular cluster of mass \(m_\text{GC}\) in the gravitational field of the Fornax dwarf galaxy subject, in addition, to the dynamical friction force due to perturbed ultralight dark matter, is governed by the equation
\begin{equation}
    m_\text{GC} \ddot{\vec{r}}=\vec{F}_\text{grav} +\vec{F}_\text{fr},
\label{equation-motion}
\end{equation}
where the dynamical friction force is given by Eq.~(\ref{dynamical-force-sphere-total-tangential-1})
and the gravitational force equals
\begin{equation}
    \vec{F}_{\mathrm{grav}} = -\frac{G m_\text{GC} M(r)}{r^2} \hat{\boldsymbol{r}}.
\end{equation}
Here, the enclosed mass of the Fornax galaxy within radius \( r \) is
\begin{equation}
    M(r) = \int\limits_0^r\!\! 4\pi \rho(r')r^{\prime 2} \,dr',
\end{equation}
where \( \rho(r) \) is the ULDM density profile.

Decomposing the equation of motion (\ref{equation-motion}) into the radial and azimuthal components \( (\hat{\boldsymbol{r}}, \hat{\boldsymbol{\phi}}) \), we find
\begin{equation}
    m_\text{GC}\!\left(\ddot{r} - r\dot{\phi}^2\right)\!\hat{\boldsymbol{r}}
    + m_\text{GC}\!\left(r\ddot{\phi} + 2\dot{r}\dot{\phi}\right)\!\hat{\boldsymbol{\phi}}
    = -\frac{G m_\text{GC} M(r)}{r^2}\hat{\boldsymbol{r}} - F_r\hat{\boldsymbol{r}} - F_t\hat{\boldsymbol{\phi}}.
\end{equation}
Projecting this equation onto the radial and tangential directions gives
\begin{equation}
    \begin{cases}
        m_\text{GC}\!\left(\ddot{r} - r\dot{\phi}^2\right)
        =  -\frac{G m_\text{GC} M(r)}{r^2} - F_r, \\[0.8em]
        \dfrac{1}{r}\dfrac{d}{dt}\!\left(m_\text{GC}r^2\dot{\phi}\right)
        = -F_t.
    \end{cases}
\end{equation}

Introducing the angular momentum
\begin{equation}
    l = m_\text{GC}r^2\dot{\phi},
\end{equation}
the above equations can be rewritten in form \eqref{EqMot}, where the term $F_r(t)$  is neglected because it is much smaller than the first term.

\newpage

\bibliographystyle{JHEP}
\bibliography{bibliography}

\end{document}